\documentclass[reprint,
superscriptaddress,
amsmath,
amssymb,
aps,
nofootinbib
]{revtex4-2}

\usepackage[normalem]{ulem}
\usepackage{graphicx,setspace}
\usepackage{dcolumn}
\usepackage{bm}
\usepackage{cancel}
\usepackage{xcolor}
\usepackage{breqn}

\usepackage{braket}
\usepackage{float}
\usepackage{physics}
\usepackage[titletoc]{appendix}

\usepackage{hyperref}
\usepackage{tabularray}
\UseTblrLibrary{amsmath}

\newcommand{\qo}[1]{``#1''}   

\makeatletter
\let\cat@comma@active\@empty
\makeatother

\renewcommand{\figurename}{{Figure}}

\begin{document}
\preprint{APS/123-QED}
\title{Tunable Quantum Interference in Free Space with a Liquid-Crystal Metagrating}

\author{Maria Gorizia Ammendola}
\thanks{MGA and IMF contributed equally to this work.}
\affiliation{Dipartimento di Fisica, Università degli Studi di Napoli Federico II, Complesso Universitario di Monte Sant’Angelo, Via Cintia, 80126 Napoli, Italy}
\affiliation{Scuola Superiore Meridionale, Via Mezzocannone, 4, 80138 Napoli, Italy}

\author{Italo Machuca Flores}%
\thanks{MGA and IMF contributed equally to this work.}
\affiliation{Dipartimento di Fisica, Università degli Studi di Napoli Federico II, Complesso Universitario di Monte Sant’Angelo, Via Cintia, 80126 Napoli, Italy}

\author{Sneha Dey}%
\affiliation{Dipartimento di Fisica, Università degli Studi di Napoli Federico II, Complesso Universitario di Monte Sant’Angelo, Via Cintia, 80126 Napoli, Italy}

\author{Francesco Di Colandrea}%
\affiliation{Dipartimento di Fisica, Università degli Studi di Napoli Federico II, Complesso Universitario di Monte Sant’Angelo, Via Cintia, 80126 Napoli, Italy}

\author{Andrei Nomerotski}
\affiliation{Faculty of Nuclear Sciences and Physical Engineering, Czech Technical University in Prague, Břehová 7, Prague, Czech Republic}
\affiliation{Institute of Physics of the Czech Academy of Sciences, Na Slovance 1999/2, Prague, Czech Republic}
\affiliation{Department of Electrical and Computer Engineering, Florida International University, 10555 West Flagler St, Miami, U.S.A}
\author{Bereneice Sephton}
\affiliation{Dipartimento di Fisica, Università degli Studi di Napoli Federico II, Complesso Universitario di Monte Sant’Angelo, Via Cintia, 80126 Napoli, Italy}

\author{Carlo Schiano}
\affiliation{Dipartimento di Fisica, Università degli Studi di Napoli Federico II, Complesso Universitario di Monte Sant’Angelo, Via Cintia, 80126 Napoli, Italy}

\author{Corrado de Lisio}
\affiliation{Dipartimento di Fisica, Università degli Studi di Napoli Federico II, Complesso Universitario di Monte Sant’Angelo, Via Cintia, 80126 Napoli, Italy}

\author{Vincenzo D'Ambrosio}
\affiliation{Dipartimento di Fisica, Università degli Studi di Napoli Federico II, Complesso Universitario di Monte Sant’Angelo, Via Cintia, 80126 Napoli, Italy}

\author{Lorenzo Marrucci}
\affiliation{Dipartimento di Fisica, Università degli Studi di Napoli Federico II, Complesso Universitario di Monte Sant’Angelo, Via Cintia, 80126 Napoli, Italy}
\affiliation{CNR-ISASI, Institute of Applied Science and Intelligent Systems, Via Campi Flegrei 34, 80078 Pozzuoli (NA), Italy}

\author{Patrick Cameron}  \email[Corresponding author: ]{p.cameron@ssmerdionale.it}
\affiliation{Dipartimento di Fisica, Università degli Studi di Napoli Federico II, Complesso Universitario di Monte Sant’Angelo, Via Cintia, 80126 Napoli, Italy}
\affiliation{Scuola Superiore Meridionale, Via Mezzocannone, 4, 80138 Napoli, Italy}

\author{Filippo Cardano}\email[Corresponding author: ]{filippo.cardano2@unina.it}
\affiliation{Dipartimento di Fisica, Università degli Studi di Napoli Federico II, Complesso Universitario di Monte Sant’Angelo, Via Cintia, 80126 Napoli, Italy}


\begin{abstract}
Structured optical materials provide a promising platform for photonic quantum information processing in free space. Beam splitters, a fundamental building block of photonic circuits, have recently been demonstrated in free space using geometric-phase optical elements. These devices coherently mix circularly-polarized transverse modes of freely-propagating optical fields, including modes carrying orbital angular momentum.
In this work, we investigate liquid-crystal metagratings as electrically tunable beam splitters for transverse-momentum optical modes. By exploiting the voltage-controlled birefringence of liquid-crystal metasurfaces, we experimentally tune the splitting ratio of the device and thereby control the degree of two-photon interference between indistinguishable photons. At the output, photons are spatially resolved on different regions of a time-resolved single-photon-sensitive detector, enabling the reconstruction of coincidence maps in the Fourier plane. This approach is readily scalable and enables highly parallel coincidence measurements across a large number of optical modes.
\end{abstract}

\maketitle 
\section{Introduction}


Interference between particles is a fundamental phenomenon that underlies many quantum technologies. It is especially relevant in the field of photonics, where direct photon-photon interaction is out of reach. Along with single-photon sources and detectors, and the use of ancillary photons, it is key for implementing multi-qubit gates, which are essential for universal quantum computing, with only linear optical components~\cite{knill_scheme_2001,obrien_optical_2007a}. In the simplest case, multi-photon, multi-mode interference manifests itself in the Hong-Ou-Mandel (HOM) effect~\cite{hong_measurement_1987}, where two photons are incident, one on each input port of a 50:50 beam splitter (BS). If these two photons are indistinguishable in all their degrees of freedom, except for that associated with the input ports, they will always exit together from either one or the other of the beam splitter's output ports. When extended to more modes and more photons, this interference gives rise to complex states that quickly become unfeasible to compute classically, such as in the famous boson sampling problem~\cite{aaronson_computational_2014a}.

To practically implement a desired quantum circuit or multi-mode interferometer, one may rely on the well-known result in quantum optics that any unitary operation may be decomposed into a connected array of only beam splitters and phase-shifters~\cite{reck_experimental_1994a}. These may be implemented using bulk optics, but such setups quickly become unfeasible as the number of modes involved increases. This has led to the development of other implementations. On-chip waveguide arrays~\cite{wang_integrated_2020} are perhaps the most developed platform for such circuits, with demonstrations of reconfigurable circuits involving up to 24 modes~\cite{barzaghi_lowloss_2025}. Alternative approaches to interferometric setups are based on direct manipulation of optical spatial modes in free space \cite{Goel2025}. These are implemented practically either via diffractive systems known as multi-plane light converters \cite{Labroille2014, Brandt2020, hiekkamaki_highdimensional_2021a}, or via combinations of spatial light modulators and high-capacity mode mixers, such as multi-mode fibers \cite{Leedumrongwatthanakun2020, Goel2024}.

Recently, the toolbox of alternative setups has been enriched by platforms based on small numbers of optical metasurfaces. These are thin materials made of patterned sub-wavelength structures~\cite{yu_light_2011}. The geometrical properties of such artificial structures allow point-wise control over the amplitude, phase, and polarization of light~\cite{chen_review_2016}. This capability has made them attractive for quantum photonics applications, leading to a recent rise in research activity in this field. Metasurfaces can couple many sets of optical modes simultaneously, making them useful for complex quantum interference experiments. For example, Liu et al.~\cite{liu_parallel_2024} use a phase-gradient metasurface to do parallel beam splitting, giving rise to new HOM interference effects and, recently, Yousef et~al. introduced a framework for generalized HOM interference with more complex metasurfaces containing multiple spatial frequency components~\cite{yousef_metasurface_2025}. Although interesting in itself, the capability for custom multi-mode interference also makes them promising for complex quantum state generation~\cite{li_metalensarraybased_2020,santiago-cruz_resonant_2022}, state manipulation~\cite{georgi_metasurface_2019,zhang_alloptical_2022}, multi-photon state measurement~\cite{wang_characterization_2023,an_efficient_2024}, and recently even multi-qubit quantum gates~\cite{xun_ultrathin_2025}.
\begin{figure*}[ht!]
\centering
\includegraphics[width=\linewidth,  trim = 0.25cm 5.8cm 0.25cm 0.15cm, clip]{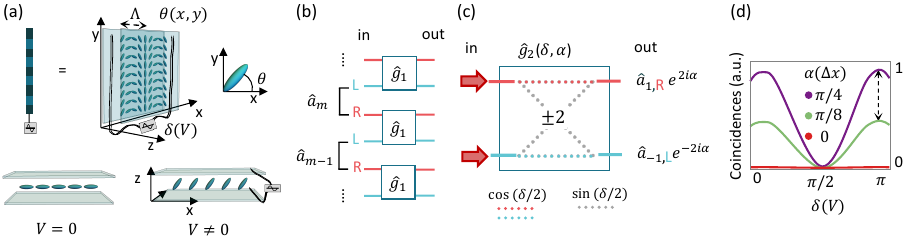}
    \caption{\textbf{Quantum interference with LCMSs.} (a) Illustration of a LCMS with the in-plane optic-axis orientation $\theta(x,y)$ patterned as a linear grating,  known as a $g$-plate. A voltage $V$ applied across the LC cell allows for tuning of the out-of-plane orientation of the LC molecules, and therefore the global birefringence parameter $\delta$ of the device. (b) A $g$-plate $\hat{g}_N$ with spatial period $\Lambda/N$ couples photons in rail $(m,L)$ with  those in rail $(m+N,R)$, and vice versa. The case $N=1$ is illustrated in this panel, with the grating coupling adjacent rails as in standard interferometric meshes. 
    (c) We zoom into the action of $\hat{g}_2(\delta)$. When $m=1$, it couples $\hat{a}_{1,R}$ and $\hat{a}_{-1,L}$ with a coupling strength (dotted lines) controlled by birefringence $\delta$. (d)~Schematic representation of HOM-like curves in coincidence counts between photons in modes $\hat{a}_{1,R}$ and $\hat{a}_{-1,L}$, when the NOON state in Eq.\ \eqref{eq:g1_lin} passes through $\hat{g}_2(\delta,\alpha)$.}
    \label{fig:concept1}
\end{figure*}
Dielectric metasurfaces can be made to perform arbitrary spatially-varying polarization transformations on optical fields \cite{PhysRevLett.118.113901}. Their elementary constituents, referred to as meta-atoms, can be accurately patterned with nanometric resolution. However, they are mostly static, their fabrication requires complex nano-fabrication methods, and there exists no analytical link between the geometrical parameters (i.e. height, shape, orientation) of meta-atoms and the elements of the local Jones matrix that models their optical action~\cite{Mueller2017}. So-called liquid-crystal metasurfaces (LCMSs) may address some of these limitations. Unlike conventional metasurfaces combined with an additional homogeneous liquid-crystal (LC) layer \cite{Petronella2025}, LCMSs consist of a thin nematic LC film in which the orientation of the molecular axes is patterned with micrometric resolution (see Fig.~\ref{fig:concept1}(a)). In the language of metasurfaces, the elementary units are the LC molecules themselves, whose only externally-controllable parameter is their orientation. As a result, a single LCMS cannot implement an arbitrary spatially-varying polarization transformation, and three cascaded layers are required \cite{dicolandrea_ultralong_2023, ammendola_largescale_2024}.
At the same time, applying a voltage across the cell enables continuous tuning of the overall optical retardance, a parameter that is fixed in standard metasurfaces. Moreover, LCMS patterning relies on cost-effective photo-alignment techniques, which significantly simplify fabrication compared to conventional metasurfaces \cite{Rubano2019}.


In this work, we demonstrate tunable quantum interference using a LC metagrating, acting as a BS for structured light. We build on the concept originally introduced by D’Ambrosio et al.~\cite{dambrosio_tunable_2019a} for vector-vortex modes of light, and extend it to spatial modes carrying quantized linear transverse momentum, rather than orbital angular momentum. 
We demonstrate control over the degree of quantum interference between pairs of incident photons by tuning the voltage applied across the LC metagrating, which controls the effective splitting ratio. 
By working with circularly-polarized transverse-momentum modes, we place our results in direct correspondence with the existing body of work on metasurface-based quantum-optics experiments~\cite{xun_ultrathin_2025, wang_metasurfaces_2022, solntsev_metasurfaces_2021, yousef_metasurface_2025}. Unlike other implementations, in our approach, these modes copropagate within a single beam, enabling a simpler and more stable experimental configuration. A Fourier lens is used to resolve the modes, which are separated into distinct spots and subsequently imaged onto a single-photon-sensitive time-stamping camera. This architecture is readily scalable and enables highly parallel coincidence measurements across a large number of modes.

\section{Concept}
\begin{figure}[ht!]
\centering
\includegraphics[width=\linewidth, trim = 0.25cm 2.9cm 8.6cm 0.25cm, clip]{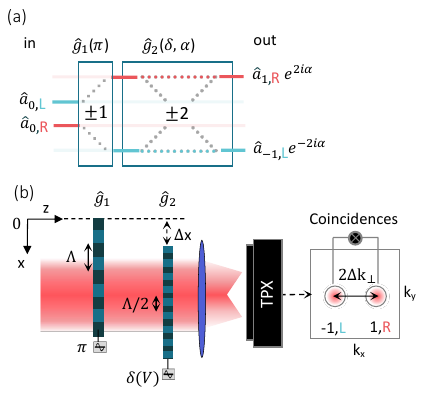}
    \caption{\textbf{Conceptual scheme of the experiment.} (a)-(b)~Two photons in the spatial mode $m=0$ pass through $\hat{g}_{1}(\pi)$ that distribute them across modes with $m=\pm1$. The transverse position corresponding to $\alpha=0$ defines the origin of the reference frame $x=0$. A second $g$-plate $\hat{g}_{2}(\delta,\alpha)$ follows and acts as a tunable BS. Both parameters $(\delta,\alpha)$ are adjustable: $\delta$ is electrically-tunable via the voltage $V$ and $\alpha$ is mechanically-tunable via a lateral shift $\Delta x$. State preparation ($\hat{g}_{1}$) and mode mixing ($\hat{g}_{2}$) occur collinearly, keeping the second device in the near field of the first. Photons are collected by a single-photon-sensitive camera (TPX) placed in the far field of a lens, where the output modes are imaged into two separated regions of interest (ROIs). Coincidence counts between events recorded in the latter regions of the sensor are reconstructed via post-processing.}
    \label{fig:concept2}
\end{figure}

This work is based on the ability of LCMSs to couple spatial and polarization modes in a controllable manner. The effect of a general LCMS can be described by a spatially-varying Jones matrix of the form
\begin{align}
Q_\delta(\theta) =
\begingroup
\renewcommand{\arraystretch}{1.5}
\begin{bmatrix}
    \cos\left(\frac{\delta}{2}\right) & i\sin\left(\frac{\delta}{2}\right)\mathrm{e}^{-2i\theta(x,y)} \\
    i\sin\left(\frac{\delta}{2}\right)\mathrm{e}^{2i\theta(x,y)} & \cos\left(\frac{\delta}{2}\right)
\end{bmatrix},
\endgroup
\label{Jonesmatrix}
\end{align}
where we adopt the circular polarization basis, with $\ket{L}=(1, 0)^T$ and $\ket{R}=(0, 1)^T$ respectively denoting left- and right-circularly polarized light. Fig.~\ref{fig:concept1}(a) shows a pictorial representation of this device. $\theta(x,y)$ is the in-plane orientation of the LC's optic axis at point $(x,y)$. $\delta$ corresponds to the device's birefringence, and it is related to the out-of-plane angle of the LCs, which can be adjusted by applying a voltage $V$ across the device \cite{piccirillo2010photon}. 
$\theta$ is not reconfigurable, but highly complex spatial patterns can be fabricated with photoalignment techniques \cite{Rubano2019}, as demonstrated in Refs.~\cite{dicolandrea_ultralong_2023,ammendola_largescale_2024}.

Let us consider a LC metagrating, hereafter referred to as $g$-plate in agreement with our original work \cite{derrico_twodimensional_2020a}, with optic-axis pattern
\begin{align}
    \theta(x,y) = \frac{\pi x}{\Lambda}+\alpha,
    \label{eq:theta}
\end{align}
where $\alpha$ is the optic-axis angle at a reference position ${x = 0}$ on the device, and $\Lambda$ is the spatial periodicity, as shown in Fig.~\ref{fig:concept1}(a).
In our experiments, we consider combined spatial-polarization modes of the electromagnetic field of the form
\begin{equation}\label{eqn:opticalmodes}
\braket{\vec{r}}{m,j} = A(x,y,z)\, e^{i k_z z}\, e^{i m \Delta k_{\perp} x}\ket{j},
\end{equation}
with spatial index $m \in \mathbb{Z}$ and polarization index $j \in (L,R)$. Here, $A(x,y,z)$ is a Gaussian amplitude-envelope with beam waist $w_0$, $\vec{r} = (x,y,z)$, and $\Delta k_{\perp}$ is a unit of transverse momentum.
In principle, there is no upper bound on the magnitude of $m$. However, we restrict attention to photons propagating predominantly along the $z$-direction, such that the longitudinal wavevector component satisfies $k_z \gg m \Delta k_{\perp}$,  ensuring operation within the paraxial regime. For visible light, this condition corresponds to values of $m$ up to a few hundred.

Spatial separation of these tilted beams occurs in the far field of a lens, where negligible overlap between neighbouring modes is guaranteed provided that the beam waist satisfies $w_0 \geq \Lambda$ \cite{derrico_twodimensional_2020a}.
We denote the action of a $g$-plate with spatial periodicity $\Lambda$ with operator $\hat{g}_{1}(\delta,\alpha)$, where: 
\begin{align}
    \hat{a}^{\dagger}_{m,L} & \xrightarrow{\hat{g}_{1}(\delta,\alpha)}  \cos(\delta/2) \hat{a}^{\dagger}_{m,L} + ie^{2i\alpha}\sin(\delta/2)\hat{a}^{\dagger}_{m+1,R}\label{eq:gplate_left} \\
    \hat{a}^{\dagger}_{m,R} & \xrightarrow{\hat{g}_{1}(\delta,\alpha)}
    \cos(\delta/2) \hat{a}^{\dagger}_{m,R} + ie^{-2i\alpha}\sin(\delta/2)\hat{a}^{\dagger}_{m-1,L}.\label{eq:gplate_right}
\end{align}
Here $\hat{a}^{\dagger}_{m,j}$ is the creation operator for a photon in mode $\ket{m,j}$ (in the following, we omit the $\dagger$ symbol for brevity). This action can be readily visualized by representing the modes as parallel rails in momentum space (see Fig.~\ref{fig:concept1}(b)). For each value of the spatial index $m$, we distinguish the $L$ and $R$ polarization rails.
The $g$-plate couples modes that lie on adjacent rails, featuring opposite polarization and different $m$ index. Specifically, it couples each $\hat{a}_{m,R}$ to $\hat{a}_{m-1,L}$ and vice versa (a $\pm 1$ shift). If the spatial period is reduced to $\Lambda/N$, with $N$ being an integer, the coupling occurs instead between $\hat{a}_{m,R}$ and $\hat{a}_{m-N,L}$ and vice versa (a $\pm N$ shift). The resulting transformation is therefore equivalent to an infinite array of parallel two-mode couplers, represented by the white boxes in Fig.~\ref{fig:concept1}(b) for the standard case $N=1$.
Moreover, Eqs.\ \eqref{eq:gplate_left} and \eqref{eq:gplate_right} show that the coupling strength depends on the birefringence $\delta$. An input photon has probability-amplitude $\cos(\delta/2)$ to remain in the same mode, and probability-amplitude $\sin(\delta/2)$ to be shifted into a neighbouring mode (see Fig.~\ref{fig:concept1}(c)). The direction of this shift, positive or negative, is determined by the photon's polarization. The phase-factor $\alpha$ is, by definition, related to the lateral offset $\Delta x$ of the grating relative to the $x = 0$ reference position, as discussed in the following paragraph.

We initially consider two photons prepared in the mode $m=0$, either with orthogonal polarizations (H,V) or (L,R). With a $\hat{g}_{1}$ at $\delta=\pi$ we convert them into the input states of our experiment, expressed as:
\begin{align}
\hat{a}_{0,H}\hat{a}_{0,V} & \xrightarrow{\hat{g}_{1}(\pi,0)} \hat{a}_{+}\hat{a}_{-}\label{eq:g1_lin}\\
\hat{a}_{0,L}\hat{a}_{0,R} & \xrightarrow{\hat{g}_{1}(\pi,0)} \hat{a}_{1,R}\hat{a}_{-1,L}.\label{eq:g1_LR}
\end{align}
Here, we define $\hat{a}_{\pm}=(\hat{a}_{-1,L} \pm \hat{a}_{1,R})/\sqrt{2}$ and we neglect global phase factors. Note that $\hat{a}_{+}\hat{a}_{-}= (\hat{a}^2_{-1,L} - \hat{a}^2_{1,R})/2$, where the cross terms cancel out for indistinguishable photons. The effect of input polarization is discussed in the Supplementary Material. We refer to these states as ``NOON" and ``circular" input states, respectively.

A second $g$-plate with spatial period $\Lambda/2$ then acts as an adjustable mode-coupler on these two input states. In fact, this mixes $\hat{a}_{-1,L}$ and $\hat{a}_{1,R}$ and vice versa:
\begin{align}
       \hat{a}_{1,R} & \xrightarrow{\hat{g}_{2}(\delta,\alpha)}  \cos(\delta/2) \hat{a}_{1,R} + ie^{-2i\alpha}\sin(\delta/2)\hat{a}_{-1,L}\label{eq:g2_L}\\
       \hat{a}_{-1,L} & \xrightarrow{\hat{g}_{2}(\delta,\alpha)}  \cos(\delta/2) \hat{a}_{-1,L} + ie^{2i\alpha}\sin(\delta/2)\hat{a}_{1,R}\label{eq:g2_R}
       . 
\end{align}
This action is depicted in Fig.~\ref{fig:concept1}(c). Moreover, the latter equations reveal that the $g$-plate $\hat{g}_{2}$ behaves as a partially polarizing BS, whose transmittance and reflectance can be continuously manipulated by tuning $\delta(V)$ and $\alpha(\Delta x)$.
This leads to a controllable HOM interference, as pictorially shown in Fig.~\ref{fig:concept1}(d) and demonstrated later in the experimental results. The exact role of $\delta$ and $\alpha$ can be found by computing the coincidence probability $P_c$ from Equations~\eqref{eq:g2_L}-\eqref{eq:g2_R}. For the circular input $\hat{a}_{1,R}\hat{a}_{-1,L}$, we find 
\begin{equation}\label{eq:Pc1}
    P_c(\delta,\alpha)=\cos^2(\delta),
\end{equation}
while for the NOON input $\hat{a}_{+}\hat{a}_{-}$,
\begin{equation}\label{eq:Pc2}
    P_c(\delta,\alpha)= \sin^2(2\alpha)  \sin^2(\delta).
\end{equation}
A complete derivation of these expressions can be found in the Methods, together with their generalization to partially distinguishable photons. Notice that, analogous to the azimuthal offset for $q$-plates~\cite{dambrosio_tunable_2019a}, interference between circularly polarized input photons has no dependence on the lateral offset $\alpha$ (Eq.~\eqref{eq:Pc1}). On the other hand, interference between linearly polarized input photons is modulated by $\alpha$ (Eq.~\eqref{eq:Pc2}). This behavior is periodic in $\alpha$ with period $\pi/2$, and in $\delta$ with period $\pi$. However, since $\sin^2{(2\alpha)}$ is always positive, it will take on recurring values after $\pi/4$. We see that, when $\alpha=0$, no interference occurs. The physical reason for this is that linearly polarized input photons $\hat{a}_{+}\hat{a}_{-}$ are eigenstates of the $\hat{g}_{2}(\delta,\alpha=0)$ transformation; as a consequence,
these two states are never mixed, regardless of $\delta$, so no interference can occur. Further discussion follows in Supplementary Material.
\begin{figure*}[ht]
\centering
\includegraphics[width=\linewidth,  trim = 0.25cm 0.25cm 0.27cm 0.25cm, clip]{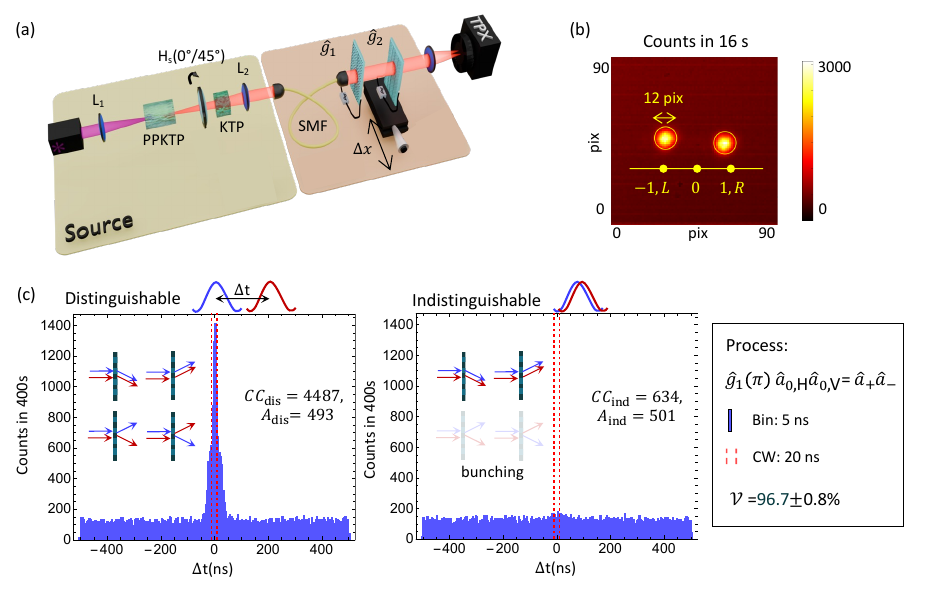}
\caption{\textbf{Experimental setup and source visibility with transverse modes.}
(a) Photons pairs are generated via type-II SPDC, focusing the pump beam on a PPKTP crystal ($L_1$). Then a half-wave plate (H$_{\text{s}}$) and a half-long KTP crystal are used to control the temporal delay between the two photons. These are recollimated with a second lens ($L_2$) and then coupled into a single mode fiber (SMF) through and sent to the tunable interference block. This stage is composed of $\hat{g}_{1}$, preparing the input state, and $\hat{g}_{2}(\delta)$, mounted on a translation stage to tune $\alpha$. Photons are then collected at the intensifier input window, placed in the far field of a lens. (b) Experimental reconstruction of counts in the two selected ROIs on the sensor, relative to the $\hat{a}_{-1,L}$ and $\hat{a}_{1,R}$ modes, performed via the post-processing algorithm described in Supplementary Material. (c) Histograms of time-of-arrival differences $\Delta t$ between photons collected in the two ROIs. H$_{\text{s}}$ at 0\textdegree ~sets zero temporal overlap (left), while at  45\textdegree  ~sets maximal temporal overlap (right), corresponding to perfect indistinguishability. The inset contains information about the process and data analysis. $\hat{g}_{1}(\pi)$ acts as a PBS on linear input photons in the $m=0$ transverse mode, showing photon bunching (no coincidences) only in the indistinguishable case. Data is collected for $400\, \text{s}$. Coincidences (CC) and average accidentals (A) are computed in the two cases, considering a CW of $20\, \text{ns}$, corresponding to $|\Delta t| \le 10\,\mathrm{ns}$. Uncertainties are computed considering Poisson-distributed variables. Error bars are too small to be visualized.}
\label{fig:expsetup}
\end{figure*}
\section{Experimental Setup}\label{sec:setup}
Fig.~\ref{fig:concept2} shows a circuit-style scheme of the experiment (with modes represented as rails), accompanied by a simplified layout of the actual experimental setup. Two orthogonal (either linear or circular) input photons are in the $m=0$ spatial mode. 
A first $g$-plate with $\delta=\pi$ prepares the input state (Eqs.~\eqref{eq:g1_lin}-\eqref{eq:g1_LR}). Its spatial period is $\Lambda=5\, \text{mm}$. As such, the beam waist is set to $w_0 \simeq5\, \text{mm}$~\cite{derrico_twodimensional_2020a} to keep spatial modes orthogonal. The $g$-plate action brings the input photons into modes $\hat{a}_{1,R}$ and $\hat{a}_{-1,L}$. This $g$-plate, labelled as $\hat{g}_{1}(\pi,0)$, defines our origin on the $x$-axis, chosen as the position where $\alpha=0$.
Then a second $g$-plate, $\hat{g}_{2}(\delta,\alpha)$, enables quantum interference between these modes (see Eqs.~\eqref{eq:g2_L}-\eqref{eq:Pc2}). The plate is mounted on a translation stage, that is used to adjust the mechanical shift $\Delta x$ relative to $x=0$. The $x=0$ value is set by exploiting near-field polarization projections, as outlined in the Supplementary Material. 
According to Eq.~\eqref{eq:theta}, the value of the residual orientation angle at the origin is $\alpha=-2\pi\Delta x/\Lambda$ (the factor 2 is due to the spatial period of this plate being $\Lambda/2$). The two-photon interference at $\hat{g}_2$ output ports  can be tuned by adjusting the voltage $V$ and the transverse position $\Delta x$.

Transverse-momentum modes associated with these output ports are spatially separated into an array of spots with a single Fourier lens. 
We detect photons with an event-based camera (TPX) with a nominal temporal resolution of $1.56\, \text{ns}$~\cite{nomerotski2019} and an effective temporal resolution of around $6\, \text{ns}$~\cite{Nomerotski2023} due to the jitter introduced by the intensifier unit of the camera. 
The different regions of interest (ROIs) where the separated spots are incident on the sensor effectively behave as separate photodetectors, enabling parallelizable coincidence detection. The possibility of simple spatial sorting of the output modes, combined with the ability to detect many modes simultaneously, is a major benefit of using these modes over, for example, OAM modes, which require either sequential projective measurements or a dedicated mode sorter. We measure the number of coincidences --- two single-photon detection events occurring within a given coincidence time window (CW) --- between pairs of output modes (in this case $\hat{a}_{1,R}$ and $\hat{a}_{-1,L}$, shown in Fig.~\ref{fig:concept2}(b)).
The post-processing procedure to reconstruct coincidence counts between different ROIs is outlined in the Supplementary Material.

A scheme of the source and of the main components of the complete experimental setup is shown in Fig.~\ref{fig:expsetup}(a). Photon pairs are generated via spontaneous parametric down conversion (SPDC).
A horizontally-polarized, $405\, \text{nm}$ continuous wave laser with a power of $2\, \text{mW}$ is focused onto a  $10$-mm-long, periodically poled potassium titanyl phosphate (PPKTP) crystal cut for type II, collinear phase matching. The temperature of the crystal is adjusted so that pairs of orthogonally polarized (horizontal and vertical) photons, referred to as signal and idler, are generated with their spectral distribution centered around  $810\, \text{nm}$ (degenerate configuration). Propagation through the PPKTP crystal itself introduces a relative temporal delay between signal and idler photons, as their polarization direction matches the fast and slow axes of the crystal. This is compensated by placing a $5$-mm-long, potassium titanyl phosphate (KTP) crystal after the PPKTP~\cite{KwiattypeIISPDC}, oriented such that their optic axes are aligned. A half-wave plate (HWP) H$_{\text{s}}$, at $45^\circ$, is placed between them to rotate the polarization of the photons by $90^\circ$, so as to switch the effective optical paths and compensate the delay. Alternatively, setting H$_{\text{s}}$ to $0^\circ$ increases the temporal delay between photons, thereby increasing their distinguishability.

After the second crystal, a dichroic filter and a low-pass filter remove the pump laser light, while a narrow band-pass filter centred at $810\, \text{nm}$ (FWHM $3\, \text{nm}$) selects the (approximately) spectrally indistinguishable photon pairs. A second lens collimates the photon pairs which are then, using a triplet collimator, coupled into a single-mode fiber (SMF). This is used to spatially filter them, selecting only the Gaussian $m=0$ mode.
After the SMF,  the pairs are out-coupled with a second triplet collimator, and a HWP and a quarter-wave plate (QWP) (not shown in the figure) are used to perform polarization compensation and prepare the state $\hat{a}_{0,H}\hat{a}_{0,V}$ at the fiber output.

\section{Results}\label{sec:results}
By directly sending the state in Eq.\ \eqref{eq:g1_LR} to the TPX camera, we can quantify the interference visibility of the photons, as in standard HOM experiments. Indeed, photon bunching in modes $\hat{a}_{1,R}$ and $\hat{a}_{-1,L}$ (Fig.~\ref{fig:expsetup}(b)) takes place at the exit of $\hat{g}_{1}(\pi,0)$ only for indistinguishable photons. Fig.~\ref{fig:expsetup}(c) shows the histogram of time differences ($\Delta t$) between photons collected in these two modes by the TPX sensor in the case of distinguishable and indistinguishable photons. As expected, few coincidences are found in the indistinguishable case compared to the case of distinguishable photons. The corresponding visibility is $\mathcal{V}=1-(CC_{ind}-A_{ind})/(CC_{dis}-A_{dis})=96.7\pm0.8\%$, where $CC$ and $A$ are the number of measured coincidences and average accidentals in the two cases, respectively. Here, we set $CW=20\, \text{ns}$ and an exposure time of $400\, \text{s}$. We benchmark these results with those obtained when detecting photons with avalanche photodiodes (APD) and separating them with a standard polarizing BS. In this case, we obtain $\mathcal{V}_{APD}=96.56\pm 0.11\%$. These visibility values are compatible within the experimental uncertainties, with the latter being higher for the TPX case, due to the TPX higher jitter and lower detection efficiency. Further details about the characterization of the source visibility with APDs, the quantitative assessment of the accidentals, and the TPX efficiency are provided in the Supplementary Material.

\begin{figure}[h!]
    \centering
    \includegraphics[width=\linewidth,  trim = 0cm 0.9cm 8.6cm 0.1cm, clip]{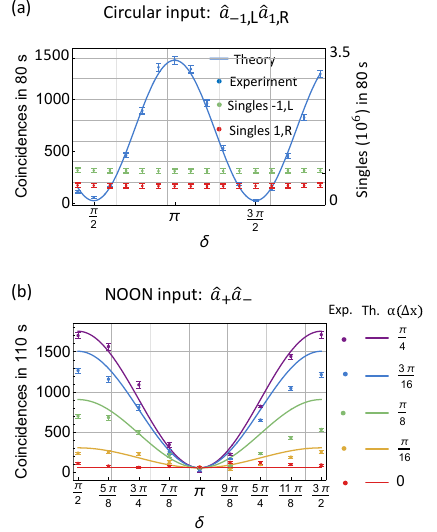}
    \caption{\textbf{Dependence of HOM interference on $\delta$ and $\alpha$} (a) Photons are injected separately in the two circular input modes $\hat{a}_{-1,L}\hat{a}_{1,R}$. Tuning $\hat{g}_{2}(\delta)$, quantum interference occurs for any value of $\alpha$ in coincidence counts accumulated for $80\, \text{s}$ (Experiment), in agreement with the theoretical model (Theory), while the singles counts in the two modes stays constant. (b) When photons are prepared in the $\hat{a}_{-}\hat{a}_{+}$ NOON input state, different degrees of quantum interference can be obtained by changing $\alpha(\Delta x)$, with $\Delta x$  the relative shift between the two $g$-plates. Coincidences are collected for $110\, \text{s}$. Experimental points follow the theoretical curves. Error bars are extracted from Poissonian statistics.}
    \label{fig:resultsHOM}
\end{figure}

After quantifying the indistinguishability of our photons, we set H$_{\text{s}}$ at $45^\circ$ (corresponding to indistinguishable photons) and tune $\hat{g}_{2}$ to demonstrate Eqs.~\eqref{eq:Pc1} and~\eqref{eq:Pc2}. Fig.~\ref{fig:resultsHOM}(a) shows the effect of tuning the voltage $V$. For the circular input, the interference curve of coincidences measured in $80\, \text{s}$ follows the theoretical behavior $P_c(\delta)$, as predicted by Eq.~\eqref{eq:Pc1}, with no dependence on $\alpha$. Here, we vary $\delta$ in the range $[\pi/2-\pi/10,2\pi+\pi/10]$, with steps of $\pi/10$. Single-photon counts in the two modes are constant, proving that any change in the coincidence rate is not due to a change of photon flux in the two states. Setting a NOON input allows us to also explore the effect of the relative position $\Delta x$ between the two $g$-plates on the two-photon interference.
Fig.~\ref{fig:resultsHOM}(b) shows five interference curves, each one corresponding to a different value of $\alpha \in [0,\pi/4]$ with steps of $\pi/16$, corresponding to $\Delta x \in [0, -\Lambda/8]=[0, -0.63\, \text{mm}]$, with steps of $\Lambda/32=0.16\, \text{mm}$. For each curve, we vary the voltage $V$ such that $\delta(V) \in [\pi/2,3\pi/2]$, with steps of $\pi/8$.

For both input states, the theoretical curves are obtained by multiplying the theoretical probability $P_c$ for partially distinguishable photons by the maximum value of coincidences in the experiment. We use the value $\mathcal{V}$ to model partial distinguishability of photons --see Methods for the derivation of Eqs.~\eqref{eq:Pc1} and~\eqref{eq:Pc2} in this case. Error bars are computed taking into account the Poissonian nature of the measured coincidences counts, as outlined in the Supplementary Material.
It is worth noting that, differently from Ref.~\cite{dambrosio_tunable_2019a}, here we demonstrate several intermediate regimes between $\alpha=0$ and $\alpha=\pi/4$, with the detection stage simply implemented using a camera-like sensor instead of coupling structured photons into SMFs.

\section{Conclusion}
In this work, we demonstrate tunable quantum interference by exploiting the electrical reconfigurability of liquid-crystal metasurfaces and the mechanical displacements between them. To demonstrate this concept, we measure the Hong-Ou-Mandel interference between two indistinguishable photons, by using a LC metagrating that acts as a tunable beam splitter for structured light in free space. By changing the voltage across the device, we effectively control the transmittance of such a \qo{beam splitter} in the mode space, which allows for tuning of photon bunching at the output. We also explore experimentally the role of the lateral offset of the LCMSs, which provides another degree of freedom to control the interference. 

In this paper, we show a relatively simple case considering interference between only two spatial optical modes.  In this case, with two modes, it would be simple to use two APD detectors and time-tagging electronics to perform the coincidence measurement; however, this lacks scalability. The goal of this paper is to lay the groundwork for more complex many-mode experiments, thanks to the source configuration, which generates copropagating photons, and the parallel detection scheme enabled by the TPX camera. Here, the $g$-plate could be simply replaced with another, more complex LCMS, generating more output modes resulting in more spots on the camera. 
The only requirement would be to adapt the data analysis, without the need for installing and aligning additional detectors and equipment.

This platform requires two $g$-plates in the near field to prepare the input state and make the two photons interfere. To avoid the effects of diffraction, the total propagation distance through the setup must be kept below the Rayleigh range. In this work, we have a Rayleigh range of $z_R \approx  100$ m, corresponding to a maximum transverse kick of $|m|\approx 150$ before breaking the paraxial approximation. Accordingly, diffraction between plates was far from being a limiting factor in the present experiment. Our scheme is thus well-positioned to take advantage of the emerging topic of metasurfaces for high-dimensional quantum optics in free space. 


\section*{Acknowledgements}
We acknowledge Yingwen Zhang for fruitful discussions. This work was supported by the PNRR MUR project PE0000023-NQSTI and PNRR
MUR project CN 00000013—ICSC.
\section*{Methods}
\subsection*{Derivation of Eqs.~\eqref{eq:Pc1}-\eqref{eq:Pc2}}
The exact role of $\delta$ and $\alpha$ in the coincidence probability $P_c$ can be found from Eqs.~\eqref{eq:g2_L}-\eqref{eq:g2_R}, defining the action of $\hat{g}_{2}(\delta,\alpha)$. For the circular input, we find
\begin{align*}
    \hat{a}_{-1,L}\hat{a}_{1,R}&\xrightarrow{\hat{g}_{2}(\delta,\alpha)}\\
    &\cos{\delta}\hat{a}_{-1,L}\hat{a}_{1,R} + \frac{i}{2} \sin{\delta}(e^{-2i\alpha}\hat{a}_{-1,L}^2+e^{2i\alpha}\hat{a}_{1,R}^2).
\end{align*}
This leads to a probability of simultaneously collecting photons in $\hat{a}_{-1,L}$ and $\hat{a}_{1,R}$:
\begin{equation}
    P_c(\delta,\alpha)=\cos^2(\delta).
\end{equation}
For the linear input, it is useful to express the output in the $\hat{a}_{-1,L}\hat{a}_{1,R}$ basis, where we find:
\begin{align*}
\hat{a}_{-}\hat{a}_{+}&\xrightarrow{\hat{g}_{2}(\delta,\alpha)}\sin{2\alpha}\sin{\delta}\hat{a}_{-1,L}\hat{a}_{1,R}+\\
   &+\frac{1}{2}\big[e^{2i\alpha}\hat{a}^2_{1,R}(\cos{2\alpha}-i\cos{\delta}\sin{2\alpha})+\\
   &-e^{-2i\alpha}\hat{a}^2_{-1,L}(\cos{2\alpha}+i\cos{\delta}\sin{2\alpha}) \big].
\end{align*}
This leads to a probability of simultaneously collecting photons in $\hat{a}_{-1,L}$ and $\hat{a}_{1,R}$:
\begin{equation}
    P_c(\delta,\alpha)= \sin^2{(2\alpha)}  \sin^2{(\delta)}.
\end{equation}
To model imperfections of the source, all theoretical curves present in the paper take into account the visibility of the source $\mathcal{V}=96.7 \pm 0.8\%$, characterized in Fig.~\ref{fig:expsetup}(c), where:
\begin{equation}
    P_c=\mathcal{V} P_{i} + (1 - \mathcal{V})P_{d},
\end{equation}
and $P_i$ and $P_d$ are the coincidence probability for two fully indistinguishable and fully distinguishable photons, respectively~\cite{tichy2014interference}.
From Eqs.~\eqref{eq:g2_L}-\eqref{eq:g2_R}, it is easy to find that, for a circular input $\hat{b}_{-1,L}\hat{a}_{1,R}$:
\begin{equation}
    P_d(\delta,\alpha=0)=\cos^4{(\delta/2)}+\sin^4{(\delta/2)},
\end{equation}
while for a linear input $\hat{b}_{+}\hat{a}_{-}$:
\begin{equation}
    P_d=\frac{1}{4}[(1+\sin{2\alpha}\sin{\delta})^2+(\sin{2\alpha}\sin{\delta}-1)^2],
\end{equation}
where we used the label $b$ to distinguish the first photon from the second. 
\bibliographystyle{apsrev4-2}
\bibliography{references}

\newpage
\clearpage
\onecolumngrid
\renewcommand{\figurename}{{Figure}}
\setcounter{figure}{0} \renewcommand{\thefigure}{{S{\arabic{figure}}}}
\setcounter{table}{0} \renewcommand{\thetable}{S\arabic{table}}
\setcounter{section}{0} \renewcommand{\thesection}{S\arabic{section}}
\setcounter{equation}{0} \renewcommand{\theequation}{S\arabic{equation}}
\onecolumngrid

\begin{center}
{\Large \textbf{Supplementary Material for:\\ Tunable Quantum Interference in Free Space with a Liquid-Crystal Metagrating}}
\end{center}

\section*{Dependence of quantum interference on the input state}
By preparing the linear state $\hat{a}_{+}\hat{a}_{-}=\hat{g}_{1}(\hat{a}_{0,H}\hat{a}_{0,V})$ and tuning the relative shift $\alpha(\Delta x)$, we can tune the level of quantum interference in $\hat{a}_{-1,L}$ and $\hat{a}_{1,R}$.
When $\alpha=0$, no interference occurs. The physical reason for this is that linearly polarized input photons $\hat{a}_{+}$ and $\hat{a}_{-}$ are eigenstates of the $\hat{g}_{2}(\delta,\alpha=0)$ transformation; as a consequence, these two states are never mixed, regardless of $\delta$, so no interference can occur.

Assume we rotate the linear input polarization to $\hat{a}_{0,\varphi/\varphi^{\perp}}=(\hat{a}_{0,R}\pm e^{i\varphi}\hat{a}_{0,L})/\sqrt{2}$, so that the input state becomes
$\hat{a}_{0,\varphi}\hat{a}_{0,\varphi^{\perp}} \xrightarrow{\hat{g}_{1}(\pi,0)} (\hat{a}^2_{-1,L} - e^{2 i \varphi}\hat{a}^2_{1,R})/2
$, which is a NOON state. The NOON states resulting from $\hat{a}_{0,H}\hat{a}_{0,V}$ and $\hat{a}_{0,D}\hat{a}_{0,A}$ inputs (corresponding to the cases $\varphi=0$ and $\varphi=\pi/2$) are $(\hat{a}^2_{-1,L} - \hat{a}^2_{1,R})/2$ and $(\hat{a}^2_{-1,L} +\hat{a}^2_{1,R})/2$, respectively.

\begin{figure}[h!]
    \centering
    \includegraphics[width=\linewidth,  trim = 0.25cm 5.5cm 0.27cm 0.25cm, clip]{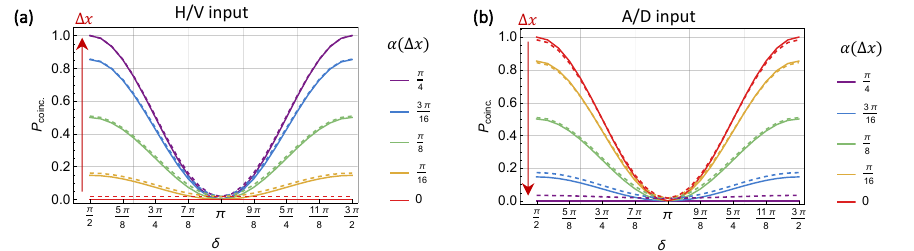}
    \caption{\textbf{Varying $\alpha$ for different linear input states.} By preparing linear input states $\hat{a}_{0,\varphi}\hat{a}_{0,\varphi^{\perp}}$ at the input of the first $g$-plate, different degrees of quantum interference can be obtained by changing $\alpha(\Delta x)$, where $\Delta x$ is the relative lateral shift between the two $g$-plates. We show the difference between the two complementary cases (a) $\hat{a}_{0,H}\hat{a}_{0,V}$, where $\varphi=0$ (which is the case we also explore experimentally in the main paper), and (b) $\hat{a}_{0,A}\hat{a}_{0,D}$, where $\varphi=\pi/2$. When $\Delta x$ is increased (red arrow), in the first case the interference emerges, while in the second case it vanishes. Plain lines refer to fully indistinguishable photons, while dashed lines correspond to partially distinguishable photons with a HOM visibility of $96.7\%$, as measured in our experiment.}\label{fig:S1}
\end{figure}

In general, the two single photon states $\hat{a}_{0,\varphi}$ and $\hat{a}_{0,\varphi^{\perp}}$ are no longer eigenstates of the $\hat{g}_{2}$ transformation for $\alpha=0$, and instead will be eigenstates of $\hat{g}_{2}$ shifted such that $\alpha=\varphi/2$.
Fig.~\ref{fig:S1} shows the difference between the obtained coincidence probability for horizontal/vertical input polarizations, $\hat{g}_{1}(\pi,0)\hat{a}_{0,H}\hat{a}_{0,V}$ (where $\varphi=0$, which is the case we also explore experimentally in Fig~\ref{fig:resultsHOM}(b)), and diagonal/antidiagonal input polarizations, $\hat{g}_{1}(\pi,0)\hat{a}_{0,D}\hat{a}_{0,A}$ (where $\varphi=\pi/2$), for the same positions of $\hat{g}_{2}$ (and thus the same values of $\alpha$). In the second case, no interference occurs when $\alpha=\pi/4$, as expected. In fact, from Eqs.~\eqref{eq:g2_L}-\eqref{eq:g2_R}, one can show that for a diagonal/antidiagonal input the resulting coincidence probability is $P_c(\delta,\alpha)= \cos^2(2\alpha)\sin^2(\delta)$. 
When $\Delta x$ is increased, in the first case the interference appears, in the second case it disappears.

Dashed curves are relative to 
distinguishable photons and are calculated with the procedure outlined in Methods~\cite{tichy2014interference}. 

\section*{Experimental reconstruction of $\delta$ and $\alpha$}
\begin{figure}[h!]
    \centering
    \includegraphics[width=\linewidth,  trim = 0.25cm 6cm 0.27cm 0.25cm, clip]{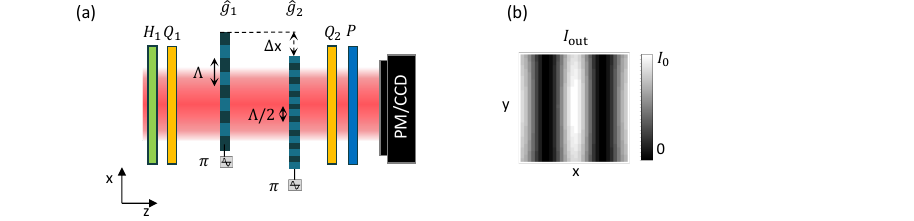}
    \caption{\textbf{Experimental characterization of $\delta$ and $\alpha$.} (a) Setup for characterization of $\delta$ and $\alpha$ via polarimetric measurements, recording light intensity with a power-meter (PM) and a CCD camera, respectively. $H_1$ and $Q_1$ are waveplates setting the input polarization, while $H_2$ and $P$ perform the projetion, as outlined in the text. (b) Simulation of $I_{\text{out}(x)}$ when $\alpha=0$. For other values of $\alpha$, fringes will move along $x$.}\label{fig:S2}
\end{figure}

In order to estimate the values of the birefringence for the two $g$-plates, that here we label $\delta_1$ and $\delta_2$, and their relative position $\Delta x$ to set $\alpha=0$, we use polarimetric measurements (setup scheme for characterization in Fig.~\ref{fig:S2}(a)). 
We use a $810$ nm superluminescent diode source and prepare a certain polarization input state $\ket{j}$ with a half-wave plate (HWP) and a quarter-wave plate (QWP) ($H_1$ and $Q_1$).
After the beam passes through the $g$-plate, we project onto an output state $\ket{j'}$ via a second QWP and a linear polarizer ($Q_2$ and $P$). A power meter was used for the determination of $\delta_1$ and $\delta_2$, while a CCD camera was used to measure $\Delta x$ and set $\alpha = 0$.
The Jones matrix of the $g$-plates reads (see Eq.~\eqref{Jonesmatrix})
\begin{align}
\hat{g}_{i}(x)=Q_{\delta_i,\alpha_i}(x) =
\begingroup
\renewcommand{\arraystretch}{1.5}
\begin{bmatrix}
    \cos\left(\frac{\delta_i}{2}\right) & i\sin\left(\frac{\delta_i}{2}\right)\mathrm{e}^{-2i(\frac{\pi}{\Lambda}x+\alpha_i))} \\
    i\sin\left(\frac{\delta_i}{2}\right)\mathrm{e}^{2i(\frac{\pi}{\Lambda}x+\alpha_i)}& \cos\left(\frac{\delta_i}{2}\right)
\end{bmatrix},
\endgroup
\end{align}
with $i\in\{1,2\}$, $\alpha_1=0$ and $\alpha_2=\alpha=-2\pi\Delta x/\Lambda$ (the factor 2 comes from the spatial period of $\hat{g}_2$, that is $\Lambda/2$).

The intensity measured by the detector at each point $x$ in the transverse plane after $\hat{g}_{1}(x)$ will be $I_{\text{out}}(x)=I_0 \abs{\bra{j'}\hat{g}_{1}(x)\ket{j}}^2$, where $I_0$ is the total input.

In particular, for a left-circular input, the difference between output intensities for opposite circular projections is:
\begin{align}
    I_L-I_R=I_0[\abs{\bra{L}\hat{g}_{1}(x)\ket{L}}^2-\abs{\bra{R}\hat{g}_{1}(x)\ket{L}}^2] = I_0 [\cos^2{(\delta_1/2)}-\sin^2{(\delta_1/2)}]= I_0 \cos{(\delta_1)}.
\end{align}
The dependence of $\delta$ on the voltage $V$ can thus be reconstructed experimentally by varying the peak-to-peak amplitude of the applied AC voltage and measuring $I_L(V)-I_R(V)$, from which we extract $\delta_{\text{exp}}(V)=\arccos\left({\frac{I_L(V)-I_R(V)}{I_0}}\right)$. This functionality of our devices physically arises from the torque exerted on the LC molecules by the applied electric field~\cite{piccirillo2010photon}, which also exhibits a strong dependence on temperature. 

In order to set the two $g$-plates with initial transverse position such that $\Delta x=0$, we use the CCD camera to measure the spatially varying polarization of the near-field beam via the projection:
\begin{equation}
    I_{\text{out}}(x)=\abs{\bra{H}\hat{g}_{2}(x)\hat{g}_{1}(x)\ket{L}}^2.
\end{equation}
In particular, we notice that, if we set either $\delta_1=\pi$ and $\delta_2=2\pi$ or $\delta_1=\delta_2=\pi$, we expect the same $I_{\text{out}}(x)$ in the two cases only when $\alpha=0$, while for other values of $\alpha$ the intensity fringes will translate along the $x$-direction. This is the criterion used for the identification of the relative position corresponding to $\alpha(\Delta x)=0$. 
Uncertainties on these measurements arise from instrumental errors, namely $\Delta_V \simeq 0.1$V for the voltage and $\Delta_x\simeq0.02$mm for the position of the second $g$-plate, controlled remotely with a motorized translation mount~\cite{Thorlabs_MTS100_2025}. Deviations may also arise from systematic errors, such as the temperature dependence of $\delta$, which we have not characterized here.

\section*{Details on source and full setup}
\begin{figure}[ht]
    \centering
    \includegraphics[width=0.98\linewidth]{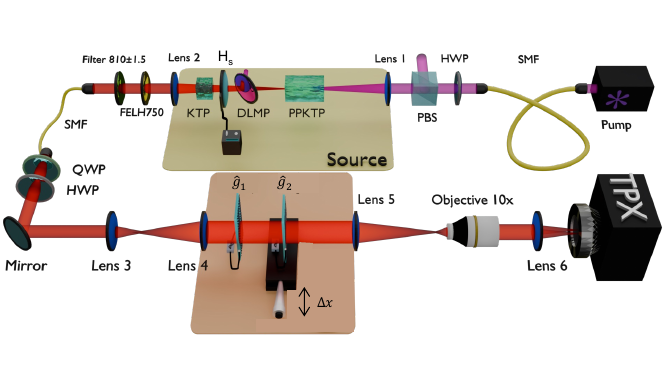}
    \caption{\textbf{Complete experimental setup.} Complete experimental setup for the creation of two-photon states, polarization control and manipulation, HOM two-mode interference, and parallel detection scheme enabled by the TPX camera.}\label{fig:S3}
\end{figure}

Fig.~\ref{fig:S3} shows the full experimental setup. The pump beam is generated by a continuous-wave laser (TopMode, TOPTICA) operating at a wavelength of $405$ nm, coupled into a single-mode fiber (SMF). The pump power and polarization are controlled using a HWP and a polarizing beam splitter (PBS). The source includes a confocal system of two lenses, both with focal length $f=200$ mm (Lens 1 and Lens 2), with a periodically poled potassium titanyl phosphate (PPKTP) crystal of size $1\times$2$\times 10 \, \text{mm}^{3}$ positioned at their focal point. The residual pump beam is removed with a dichroic mirror (DMLP). An electronic HWP (H$_{\text{s}}$) and a potassium titanyl phosphate (KTP) crystal of size $3\times$3$\times5 \, \text{mm}^{3}$ are used to compensate the relative temporal delay between the produced photons. 

The phase matching within the PPKTP crystal is temperature dependent. The crystal's temperature is precisely controlled using an oven (from Raicol), with the optimal pair-production rate found at $\simeq 32\,\text{\textdegree}$. To enforce complete removal of the pump, further filtering is achieved with a long-pass filter with a cutoff frequency of 750 nm (Thorlabs FELH750). Finally, degenerate photon pairs are selected using a $810.0\pm1.5$ nm bandpass filter (Semrock).

Photon pairs are then coupled into a single-mode fiber (SMF). After the SMF, the pairs' polarization is restored using a QWP and HWP. The beam waist after the fiber out-coupler is measured to be approximately $0.65 \, \text{mm}$, so an $8\times$ magnification system is implemented with Lens 3 ($f=50\, \text{mm}$) and Lens 4 ($f=400\, \text{mm}$) to expand the beam waist to $\simeq5.2\, \text{mm}$ at the first $g$-plate's plane, thus matching $\Lambda=5\, \text{mm}$. 

The interference stage includes two $g$-plates ($\hat{g}_{1}$ and $\hat{g}_{2}$), with the second mounted on a translation stage for precise control of $\Delta x$. Lens 5 ($f=500\, \text{mm}$) allows us to go to the far field, where the transverse momentum modes are spatially separated into spots. Due to the large sensor size of the TPX camera, these spots are magnified and imaged using am $11\times$ magnification system composed of a $10\times$ objective (with effective focal length $18\, \text{mm}$) and Lens 6 ($f=200\, \text{mm}$). The magnification factor was chosen to obtain a suitable spot size on the sensor, with the spot diameter covering about $12$ pixels. If the spot covers too few pixels, it may easily saturate the image intensifier. On the other hand, a spot spread over too many pixels would result in increased noise due to dark counts. The spot separation (which here is $\gtrsim20$ pixels) must be kept above a certain threshold due to the presence of afterpulses from the image intensifier, which results in false coincidences within a radius of approximately 8 pixels~\cite{mahon_study_2024}. Finally, the detection efficiency of the TPX (with the image intensifier) was estimated to be $\sim7\%$, following the method outlined in Ref.~\cite{vidyapin_characterisation_2023a}.

\section*{Characterization of the source visibility with Avalanche Photodiodes}
The visibility of the source was also characterized using a polarization-based HOM interferometer, not shown in Fig.~\ref{fig:S3} for simplicity. The two photons are coupled to a SMF. At the fiber exit, their polarization is rotated to diagonal and anti-diagonal, by using a QWP and a HWP, and a polarizing beam splitter (PBS) spatially separates their horizontal and vertical polarization components. Photons exiting from the PBS output ports are detected by two avalanche photodiodes (APDs from Excelitas). This configuration allows us to measure coincidences between the output ports of the PBS for both temporally distinguishable and indistinguishable photon pairs. It is straightforward to show that, for indistinguishable photons, the input state reads $\hat{a}_{0,D}\hat{a}_{0,A}=1/2(\hat{a}_{0,H}\hat{a}^2_{0,H}-\hat{a}^2_{0,V})$, hence no coincidence is expected in this case~\cite{polarizationHOM}. By counting coincidence events when sending distinguishable ($CC_{\mathrm{dis}}$) and indistinguishable ($CC_{\mathrm{ind}}$) photon pairs, we obtain the visibility as $\mathcal{V}_{APD} = 1 -CC_{\mathrm{ind}}/CC_{\mathrm{dis}}= 96.56 \pm 0.11\% $. H$_{\text{s}}$ was set to $0^{\circ}$ and $45^{\circ}$ to achieve these two regimes (as outlined in the Main). Coincidence counts were recorded using a time tagger with a $1\ \mathrm{ns}$ coincidence window.\\

\section*{Data analysis and errors estimation in photon counting}
TPX is a time-stamping readout sensor with $256 \times 256$ pixels of $55 \times  55 \, \mu \text{m}^2$. The processing electronics in each pixel records the time of arrival of events with a $1.6$ ns timing resolution and stores it as a time code in a memory inside the pixel. The information about time-over-threshold (ToT), which is related to the deposited energy in each pixel, is also stored.
In the single-photon-sensitive operation, the camera is coupled to an intensifier, a vacuum device with a photocathode, followed by a micro-channel plate, placed before the sensor, with a gain optimized to provide the best photon detection efficiency. Each measurement acquired with the TPX camera consists of 16 seconds, where binary files (.tpx), containing the list of all the events on the sensor, are generated. The events are combined into the “clusters” using the Amsterdam Scientific Instrument software \textit{Luna} and stored in .h5 files. Clusters are groups of events adjacent to each other and within a preset temporal and spatial window. Each event in a cluster should have a neighboring event separated by no more than $100\, \text{ns}$ and $15$ pixels. Moreover, only clusters with central $ToT \geq 600$ are selected and clusters containing only one event were discarded. Signal amplification and clustering led to TPX effective temporal resolution of around $6\, \text{ns}$.

Each cluster is then considered as a single-photon event on the sensor. In this experiment, two circular regions of interest (ROIs) (radius $6$ pixels, separation $33$ pixels) were chosen to generate two lists of cluster events. These lists were compared to produce histograms of difference in arrival time, $\Delta t$, between pairs of events, like the ones in Fig.~\ref{fig:expsetup}(c). The number of coincidences, $CC$, was extracted by finding all pairs with $|\Delta t|< \tau_{window}$, with $\tau_{window}=10\, \text{ns}$. We also estimate the number of accidental coincidences, $A$, which are coincidence events between two uncorrelated events that happen to occur within the coincidence window. Accidental coincidences are independent of $\Delta t$, so we can estimate them by using a \qo{shifted} coincidence window, e.g., finding pairs with $|\Delta t - t_0|<\tau^{acc}_{window}$, where $t_0$ is the shift set to $500\, \text{ns}$ to remain sufficiently far from the real coincidences peak ($\Delta t\simeq0$). To obtain a better estimate of the accidentals, we average over a larger window $\tau^{acc}_{window}=10 \tau_{window} = 100\, \text{ns}$, so that $A'=A/10$.

Error bars are calculated taking into account the Poissonian nature of the measured coincidences counts, i.e., that the probability of detecting one coincidence event is uncorrelated with prior or subsequent coincidence detections. Therefore, the error is simply the square root of the mean:
\begin{equation}
    \sigma_{cc} = \sqrt{CC}, \quad \sigma_{a}' = \sqrt{A}/10 = \sqrt{A'/10},
\end{equation}
and
\begin{equation}
    \sigma_{counts}=\sqrt{\sigma^2_{cc}+{\sigma'}^2_{a}}.
\end{equation}
The TPX effective temporal resolution of $6\, \text{ns}$, together with its higher background signal due to the signal amplification, led to a higher number of accidentals and thus to a wider errorbar in the visibility $\mathcal{V}$ with respect to $\mathcal{V}_{APD}$.
\hspace{5pt}
        

\end{document}